\documentclass[twoside,twocolumn,9pt]{article}

\usepackage{extsizes}
\usepackage[super,sort&compress,comma]{natbib} 
\usepackage[version=3]{mhchem}
\usepackage[left=1.5cm, right=1.5cm, top=1.785cm, bottom=2.0cm,headheight=12pt]{geometry}
\usepackage{balance}
\usepackage{times,mathptmx}
\usepackage{sectsty}
\usepackage{graphicx} 
\usepackage{lastpage}
\usepackage[format=plain,justification=justified,singlelinecheck=false,font={stretch=1.125,small,sf},labelfont=bf,labelsep=space]{caption}
\usepackage{float}
\usepackage{fancyhdr}
\usepackage{fnpos}
\usepackage[english]{babel}
\usepackage{array}
\usepackage{droidsans}
\usepackage{charter}
\usepackage[T1]{fontenc}
\usepackage[usenames,dvipsnames]{xcolor}
\usepackage{setspace}
\usepackage[compact]{titlesec}
\usepackage{comment}
\definecolor{cream}{RGB}{222,217,201}

\usepackage{amsmath}

\begin{document}

\pagestyle{fancy}
\thispagestyle{plain}
%\fancypagestyle{plain}{

%%%%HEADER%%%
%\fancyhead[C]{\includegraphics[width=18.5cm]{head_foot/header_bar}}
%\fancyhead[L]{\hspace{0cm}\vspace{1.5cm}\includegraphics[height=30pt]{head_foot/journal_name}}
%\fancyhead[R]{\hspace{0cm}\vspace{1.7cm}\includegraphics[height=55pt]{head_foot/RSC_LOGO_CMYK}}
%\renewcommand{\headrulewidth}{0pt}
%}
%%%END OF HEADER%%%

%%%PAGE SETUP - Please do not change any commands within this section%%%
\makeFNbottom
\makeatletter
\renewcommand\LARGE{\@setfontsize\LARGE{15pt}{17}}
\renewcommand\Large{\@setfontsize\Large{12pt}{14}}
\renewcommand\large{\@setfontsize\large{10pt}{12}}
\renewcommand\footnotesize{\@setfontsize\footnotesize{7pt}{10}}
\makeatother

\renewcommand{\thefootnote}{\fnsymbol{footnote}}
\renewcommand\footnoterule{\vspace*{1pt}% 
\color{cream}\hrule width 3.5in height 0.4pt \color{black}\vspace*{5pt}} 
\setcounter{secnumdepth}{5}

\makeatletter 
\renewcommand\@biblabel[1]{#1}            
\renewcommand\@makefntext[1]% 
{\noindent\makebox[0pt][r]{\@thefnmark\,}#1}
\makeatother 
\renewcommand{\figurename}{\small{Fig.}~}
\sectionfont{\sffamily\Large}
\subsectionfont{\normalsize}
\subsubsectionfont{\bf}
\setstretch{1.125} %In particular, please do not alter this line.
\setlength{\skip\footins}{0.8cm}
\setlength{\footnotesep}{0.25cm}
\setlength{\jot}{10pt}
\titlespacing*{\section}{0pt}{4pt}{4pt}
\titlespacing*{\subsection}{0pt}{15pt}{1pt}
%%%END OF PAGE SETUP%%%

%%%FOOTER%%%
%\fancyfoot{}
%\fancyfoot[LO,RE]{\vspace{-7.1pt}\includegraphics[height=9pt]{head_foot/LF}}
%\fancyfoot[CO]{\vspace{-7.1pt}\hspace{13.2cm}\includegraphics{head_foot/RF}}
%\fancyfoot[CE]{\vspace{-7.2pt}\hspace{-14.2cm}\includegraphics{head_foot/RF}}
%\fancyfoot[RO]{\footnotesize{\sffamily{1--\pageref{LastPage} ~\textbar  \hspace{2pt}\thepage}}}
%\fancyfoot[LE]{\footnotesize{\sffamily{\thepage~\textbar\hspace{3.45cm} 1--\pageref{LastPage}}}}
\fancyhead{}
\renewcommand{\headrulewidth}{0pt} 
\renewcommand{\footrulewidth}{0pt}
\setlength{\arrayrulewidth}{1pt}
\setlength{\columnsep}{6.5mm}
\setlength\bibsep{1pt}

\newcommand{\HP}{H$_2$O$_2$}

\newcommand{\MO}[1]{\textcolor{blue}{#1}}
%%%END OF FOOTER%%%

%%%FIGURE SETUP - please do not change any commands within this section%%%
\makeatletter 
\newlength{\figrulesep} 
\setlength{\figrulesep}{0.5\textfloatsep} 

\newcommand{\topfigrule}{\vspace*{-1pt}% 
\noindent{\color{cream}\rule[-\figrulesep]{\columnwidth}{1.5pt}} }

\newcommand{\botfigrule}{\vspace*{-2pt}% 
\noindent{\color{cream}\rule[\figrulesep]{\columnwidth}{1.5pt}} }

\newcommand{\dblfigrule}{\vspace*{-1pt}% 
\noindent{\color{cream}\rule[-\figrulesep]{\textwidth}{1.5pt}} }
%
%\makeatother
%%%%END OF FIGURE SETUP%%%
%
%%%%TITLE, AUTHORS AND ABSTRACT%%%
\twocolumn[
 \begin{@twocolumnfalse}
 \vspace{0cm}
\sffamily
\begin{tabular}{m{1.5cm} p{13.5cm} }

&\noindent\LARGE{\textbf{Diffusiophoretic design of self-spinning microgears from colloidal microswimmers}} %\Article title goes here instead of the text "This is the title"
\vspace{0.3cm} \\

&\noindent\large{Antoine Aubret,$^{\ast}$\textit{$^{a}$} and J\'{e}r\'{e}mie Palacci\textit{$^{a}$}} \vspace{0.3cm}\\%Author names go here instead of "Full name", etc.

&\noindent\normalsize{Design strategies to assemble dissipative building blocks are essential to create novel and smart materials and machines. We recently demonstrated the hierarchical self-assembly of phoretic microswimmers into self-spinning microgears and their synchronization by diffusiophoretic interactions [Aubret \textit{et al., Nature Physics}, 2018]. In this paper, we adopt a pedagogical approach and expose our strategy to  control self-assembly and build machines using phoretic phenomena. We notably introduce Highly Inclined Laminated Optical sheets microscopy (HILO) to image and quantify anisotropic and dynamic diffusiophoretic interactions, which could not be observed by standard fluorescence microscopy. The dynamics of a (haematite) photocalytic material immersed in (hydrogen peroxide) fuel under various illumination patterns is first described and  quantitatively rationalized by a model of diffusiophoresis, the migration of a colloidal particle in a concentration gradient. It is further exploited to design phototactic microswimmers, that direct towards the high intensity of light, as a result of the the torque exerted by the haematite in a light gradient on a microswimmer.  We finally demonstrate the assembly of self-spinning microgears from colloidal microswimmers by controlling dissipative diffusiophoretic interactions,  that we characterize using HILO and quantitatively compare to analytical and numerical predictions. Because the approach described hereby is generic, this works paves the way for the rational design of machines by controlling phoretic phenomena.} \\

\end{tabular}

 \end{@twocolumnfalse} \vspace{0.6cm}
]
%%%END OF TITLE, AUTHORS AND ABSTRACT%%%

%%%FONT SETUP - please do not change any commands within this section
\renewcommand*\rmdefault{bch}\normalfont\upshape
\rmfamily
\section*{}
\vspace{-1cm}

%%%FOOTNOTES%%%
%%%FOOTNOTES%%%

\footnotetext{\textit{$^{a}$Department of Physics, University of California, San Diego, USA; E-mail: aaubret@ucsd.edu}}

\section{Introduction}
The self-assembly of individual building blocks into functional units is a fundamental goal of science and engineering. 
Progress has been made in the assembly of colloidal particles into isotropic, chiral, and directional equilibrium structures\cite{Cademartiri_Bishop-NatureMaterials-2014} using DNA origami\cite{Wagenbauer_Dietz-Nature-2017,BenZion_Chaikin-Science-2017}, entropic forces\cite{Sacanna_Pine-Nature-2010}, anistropic particles\cite{Glotzer_Solomon-NatureMat-2007,Chen_Granick-Nature-2011}, or field induced interactions \cite{Bharti_Velev-Langmuir-2015,Tasci_Marr-Nature-2016,Snezhko_Aranson-NatureMat-2011,Erb_Yellen-Nature-2009}. The resulting {\it equilibrium} structures are however static, unless an external force is applied, for instance, with optical tweezers \cite{Grier-Nature-2003} or magnetic actuation \cite{Bharti_Velev-Langmuir-2015,Li_Wang-NanoLetters-2017,Han_Velev-ScienceAdvances-2017}.\\
Dissipative self-assembly that consume energy alternatively allows architectures  to be dynamical and reconfigurable \cite{Marchetti_Simha-RevModPhy-2013,Wang_Mallouk-AccountsofChemicalResearch-2015}, a desirable feature for advanced materials.
Synthetic examples of dynamics self-assembly were initially limited to macroscopic systems \cite{Grzybowski_Whitesides-Nature-2000,Grzybowski_Whitesides-Science-2002} but recent progress in colloidal science has made available self-propelled colloids (or microswimmers)\cite{Youssef_Sacanna-NatureComm-2016,Sacanna_Yi-NatureComm-2013,Aubret_Palacci-COCIS-2017,Dey_Sen-COCIS-2016}  that are self-driven and convert available free energy into motion.
They enabled the rise of Active Matter as a field and the exploration of emergent phenomena: dynamic phase transition, clustering, living crystals, or flocking, as examples of self-organization \cite{Marchetti_Simha-RevModPhy-2013,Zottl_Stark-JPCM-2016}  as well as the field-driven phase-behaviors of collective assemblies \cite{Yan_Granick-NatureMat-2016,Yan_Granick-Nature-2012}. The design of structures made from microswimmers is however missing\cite{Wang_Mallouk-AccountsofChemicalResearch-2015,Marchetti_Simha-RevModPhy-2013}, highlighting the lack of  simple design rules in dissipative self-assembly\cite{Fialkowski_Grzybowksi-JPCB-2006,England-NatureNano-2015}. Templated assembly has been used to tackle this issue and guide assembly, by using boundaries to direct swimmers \cite{Das_Ebbens-NatureComm-2015,Simmchen_Sanchez-NatureComm-2016}, resulting in rectified motion\cite{Katuri_sanchez-ACSNano-2018,Brown_Poon-SoftMatter-2016,Takagi_Zhang-SoftMatter-2014,Das_Ebbens-NatureComm-2015,Simmchen_Sanchez-NatureComm-2016} and exploited to actuate asymmetric microgears \cite{Maggi_DiLeonardo-Small-2015,Vizsnyiczai_DiLeonardo-NatureComm-2017}. However, it relies on preliminary microfabrication and show  limited flexibility or control on the assembly.
An alternate approach uses light patterns to encode information\cite{Eskandarloo_Abbaspourrad-Nanoscale-2017,Ibele_Sen-Angewandte-2009,Palacci_Pine-Science-2013,Li_Liu-AdvancedMaterials-2016,Chatuverdi_Velegol-Langmuir-2010,Maggi_DiLeonardo-NatureComm-2015}, in place of microscopic templates. It is highly flexible as the the spatiotemporal modulation of illumination tunes the velocity of particles, and subsequent density\cite{Stenhammare_Cates-Science-2016,Arlt_Poon-NatureComm-2018}, with a micron-resolution and can conveniently be switched on and off \cite{Palacci_Chaikin-PTRSLA-2014}. The superimposition of light gradients can direct the motion of particles, as for biological cells \cite{Dervaux_Brunet-NaturePhys-2017}, self-electrophoretic nanotrees\cite{Dai_Tang-NatNano-2016,Zheng_Tang-NatureComm-2017}, or photocatalytic symmetric\cite{Chen_Zhang-AdvancedMaterials-2017,He_Lin-Angewandte-2017,Li_Liu-AdvancedMaterials-2016} or Janus\cite{Lozano_Bechinger-NatureComm-2016, Chatuverdi_Velegol-Langmuir-2010} microswimmers. \\
%
%It is therefore possible to build dynamical structures, and to encode behavioral information in the swimmers through a careful design of their fuel consumption to tune their dynamical properties \cite{Kim_Fan-Nanoscale-2015,Sacanna_Yi-NatureComm-2013,Golestanian_Adjari-NewJournalofPhysics-2007,Golestanian_Adjari-PRL-2005}.\\
In this paper, we show how light patterns can be used with photocatalytic particles to control the phenomenon of diffusiophoresis, the motion of a colloidal particle along a gradient of chemicals \cite{Anderson-AnnualReviewFM-1989,Golestanian_Adjari-PRL-2005,Buttinoni_Bechinger-JPCM-2012,Howse_Golestanian_PRL-2007,Golestanian_Adjari-NewJournalofPhysics-2007,Golestanian_Adjari-PRL-2005}. It enables the design of anisotropic and dynamic diffusiophoretic interactions that guide the robust and faithful assembly of phototactic microswimers into autonomous self-spinning microgears.   We present a step by step approach starting from a simple photoactive building block and ultimately forming a complex spinning micromachine. First, we propose a simple design to engineer phototactic microswimmers using conventional colloidal particles and a photoactive haematite component.  Next, we harness the light sensitivity of those particles  to trigger their  self-assembly in a controllable and robust manner. Guided by light patterns, serving as cues, the swimmers autonomously form chiral rotors, which sustain under uniform illumination. The rotor constitutes a dissipative building block that  creates a repulsive, anisotropic potential of diffusiophoretic origin, which we characterize using Highly Inclined Laminated Optical sheets microscopy (HILO). The results  agree with analytical and numerical predictions of  a  simple model of a rotor forming a hexapolar sink of fuel. Our results demonstrate the hierarchical construction of machines from machines by taking advantage of phoretic phenomena and light to direct self-assembly. 

%We thoroughly investigate the dynamics of the individual swimmers and rotors and demonstrate that it originates from diffusiophoresis. 
\section{Response of a photocalytic material, haematite, to light gradients}
\label{sec:haematite}
\subsection{Haematite, a photocatalytic material}
The energy-transducer in our work consists in a piece of semiconducting, photocatalytic iron oxide $\alpha-$Fe$_2$O$_3$, haematite, immersed in a solution of hydrogen peroxide \HP. A diluted solution of \HP\,  spontaneously degrades in presence of a haematite catalyst: 2\HP\,$\rightarrow$ 2H$_2$O + O$_2$. In a nutshell, haematite creates electron-hole pairs following illumination by UV-blue light. While spontaneous recombination usually occurs over a scale of a few nanoseconds, holes and electrons bounded to the surface allow redox processes to occur, catalyzing the \HP\,decomposition. The overall reaction scheme is complex, involving many intermediate species such as hydroxide radicals \cite{Lin_Gurol-EST-1998}, but eventually leads to O$_2$ and H$_2$O production. 
The release of chemical energy from the degradation of \HP\, is exploited to inject energy locally and maintain the system away from equilibrium.
\begin{figure}[th!]
\centering
\includegraphics[scale=1]{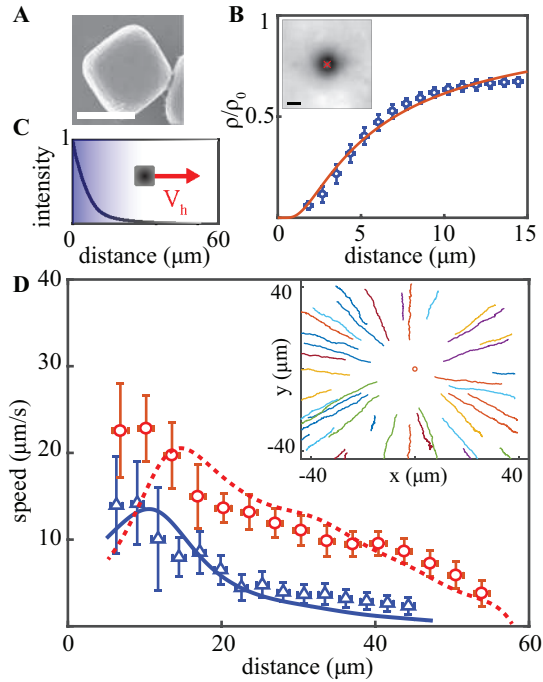}
\caption{\textbf{Response of haematite under illumination}. A) SEM image of a haematite particle. Scale bar is 500 nm. B) Isotropic phoretic repulsion of 20 nm fluorescent beads around a haematite particle tethered on the surface of a capillary, under uniform blue ($\lambda =395$ nm) excitation. The density of fluorescent beads (blue circles) decays when approaching the haematite particle, and follows eq.\eqref{eq:density_radial} (red line), with $\alpha=64$ $\mu$m$^3$/s. Inset: repulsion of FB in the XY plane. The red cross locates the haematite particle (scale bar 5 $\mu$m). C) Isotropic intensity profile generated with a slightly defocused blue laser spot ($\lambda=404$ nm), after azimuthal averaging, used for analyzing the response of haematite in a gradient of light. An asymmetric fuel consumption is generated at the surface of the haematite, triggering its motion at speed $\mathbf{V}_\text{h}$ by self-diffusiophoresis. D) Recorded speeds of haematite particles in the intensity gradient shown in C), at excitation power of P=0.16 mW (blue triangles) and P=1 mW (red circles). The velocity is adjusted with eq.\eqref{eq:haematite_speed} (solid blue line and dashed red line) from the measured intensity profile and using constant saturation intensity (see main text). Inset : trajectories of individual haematite particles, isotropically repelled from the laser spot (red dot), at P= 1mW.}
\label{fig:haematite}
\end{figure}
 We explore the response of haematite micrometric particles (size $\sim 0.5 - 1$ $\mu$m) immersed in a solution of \HP\,under various illumination patterns (Fig.\ref{fig:haematite}A, see Methods for sample preparation).

\subsection{Tethered haematite in uniform illumination}
We study the effect of fuel consumption of a single haematite under uniform illumination and examine the diffusiophoretic behavior of colloidal beads in presence of a gradient of concentration fuel. We first tether a haematite particle on the surface of a capillary and introduce a solution (pH$\sim 6$) containing 20 nm nile red fluorescent beads, which small size warrants a quick equilibration. In standard epifluorescence microscopy, the entire upper space is excited, resulting in strong background fluorescence and hindering  the observation of near-field phoretic interactions (see Methods and Fig.\ref{fig:setup}D). To overcome this problem, we use  Highly Inclined and Laminated Optical sheet microscopy technique\cite{Tokunaga_Sogawa-NatureMetehods-2008,Shashkova_Leake-BioscienceReports-2017,Liu_Betzig-MolecularCell-2015} (HILO, or near-TIRF) that narrows the observation space to a thin slice near the bottom of the capillary, where the particles sits (see Methods). In brief, the technique involves a setup similar to Total Internal Reflection Fluorescence microscopy (TIRF), with the beam forming an angle slightly smaller than the critical angle. It results in a highly inclined beam that excites only a few microns above the substrate, and allows us to analyze the density of the fluorescent beads around a haematite. In the absence of blue illumination or \HP\, fuel, the concentration of fluorescent beads is uniform. However, we an exclusion zone appears near the haematite particle when it is photo-activated by blue light: the beads are phoretically repelled and drift away from the haematite in the isotropic gradient of \HP\,consumption (Fig.\ref{fig:haematite}B).
The concentration $c$ of \HP\,  obeys the Laplace equation $\Delta c=0$ with a sink placed at the position of the haematite, $r=0$ giving   $c(r) \propto 1/r$. A colloidal particle placed in a chemical gradient exhibits a diffusiophoretic migration with a velocity $v_{DP}=D_{\text{DP}} \nabla c$, with $D_\text{DP}$ the phoretic mobility, which depends on the particle-solvent interaction\cite{Anderson-AnnualReviewFM-1989}. For the tethered haematite, we predict a phoretic repulsion for a fluorescent bead at distance $r$ with velocity $v_\text{DP}=\alpha/r^2$.
Particles conservation leads to a flux $\textbf{J}$ of density $\rho$ of beads, as the diffusiophoretic repulsion is balanced by thermal diffusion:
\begin{equation}
\label{eq:flux}
\mathbf{J}=-D_\text{c} \mathbf{\nabla} \rho + \mathbf{v_\text{DP}} \rho
\end{equation}
with $D_\text{c}=18$ $\mu$m$^2$/s the thermal diffusion coefficient of the beads. At steady state, $\mathbf{J}=0$, further integrating eq.\eqref{eq:flux}, the radial intensity profile in the 2D (XY) plane is:
\begin{equation}
\label{eq:density_radial}
\rho=\rho_0 e^{-\alpha/(D_\text{c}r)}
\end{equation}

This is analogous to an effective Boltzmann distribution of the beads in the repulsive potential $U(r) \propto c(r) = \alpha/r$.
We use eq.\eqref{eq:density_radial} to adjust our data using $\rho_0$ and $\alpha$ as fitting parameters, and find an excellent agreement for $\alpha = 64 \pm 10$ $\mu$m$^3$/s (Fig.\ref{fig:haematite}B). A haematite particle acts as a sink of \HP\,and repels the surrounding particles by diffusiophoresis.

\subsection{Untethered haematite under light gradients}
Untethered haematite in uniform light exhibit random diffusive motion.
Following, we study the behavior of untethered haematite particles in light gradients (see supplementary movie S1) by using a slightly defocused laser spot projected on the bottom surface of the sample (Fig.\ref{fig:haematite}C, $\lambda=404$ nm, see Methods). 
 We initially locate haematite particles in the sample and place the laser  in their vicinity. We then turn it on, and immediately start recording: the particles are isotropically repelled away from the laser spot. We repeat the experiment over several dozens of individual haematite particles and extract the evolution of the speed in the radial direction. The procedure is repeated for 5 different intensities, ranging from $0.03$ mW to $1$ mW (peak intensities of $\approx 0.06$ to $2$ $\mu$W/$\mu$m$^2$).
The recording is stopped once most of the particles have left the field of view. The positions are averaged over $1$s and the speed is calculated as a function of the position for each particle. We average the data over $\approx 30$ haematite particles for each intensity profile, measured and quantified independently (Fig.\ref{fig:haematite}C, see Methods).
The particles travel radially along the light gradients, migrating towards lower intensities. The speed decreases with increasing distance from the laser, as presented in Fig.\ref{fig:haematite}D. \\
As the reaction rate of decomposition of \HP\, depends on the light intensity,  the light gradient induces gradients of reaction rates at the surface of haematite, leading to a phoretic velocity $V \propto \nabla \nu$\cite{Golestanian_Adjari-NewJournalofPhysics-2007} (Fig.\ref{fig:haematite}C). The photoactivity of haematite is proportional to the absorbed intensity $I_\text{abs}$, which saturates at high intensity $I_\text{sat}^\text{h}$:  $I_\text{abs} \propto I/(1+I/I_\text{sat}^\text{h})$. We subsequently obtain the phoretic velocity for the untethered haematite particle:
\begin{equation}
\label{eq:haematite_speed}
V_{\text{h}}(I) = \chi\frac{\nabla I}{{(1+I/I_\text{sat}^\text{h})}^2}
\end{equation}
where,  $\chi$, respectively $I_\text{sat}^\text{h}$,  are adjustable parameters characterizing the phoretic response of the particle, respectively its absorption properties. The saturation intensity $I_\text{sat}^\text{h}$ is a material property, independent of the illumination intensity as the experiment is performed in a  reaction-limited regime \cite{Palacci_Chaikin-ScienceAdvances-2015}. By measuring the intensity profile $I(x)$, we compare our model with the measured velocities and observe a good agreement with eq.\eqref{eq:haematite_speed} for $I_\text{sat}^\text{h} \sim 75$nW/$\mu$m$^2$ (see Fig.\ref{fig:haematite}D). The agreement holds over 2 orders of magnitude for the illumination, ranging from $0.03$ to $1$ mW without altering $I_\text{sat}^\text{h}$.

\section{Design of  a phototactic swimmer}
\label{sec:swimmers}
We take advantage of the response of haematite to light gradients to design phototactic Janus microswimmers, consisting  of an inert polymer bead (3-(trimethoxysilyl)propyl
methacrylate, TPM, Sigma Aldrich), from which a haematite particle is extruded (see Fig.\ref{fig:swimmers}A)\cite{Youssef_Sacanna-NatureComm-2016,Sacanna_Yi-NatureComm-2013}, forming a swimmer with fore-aft asymmetry and typical length $l = 1.8$ $\mu$m. We introduce the swimmers in a capillary containing a $6\%$ solution of hydrogen peroxide, they sediment close to the surface and the observations are performed   in this plane.
In the absence of illumination, the particles undergo thermal motion and do not self-propel. 

\subsection{Microswimmer in uniform light}
\begin{figure}[th!]
\centering
\includegraphics[scale=1]{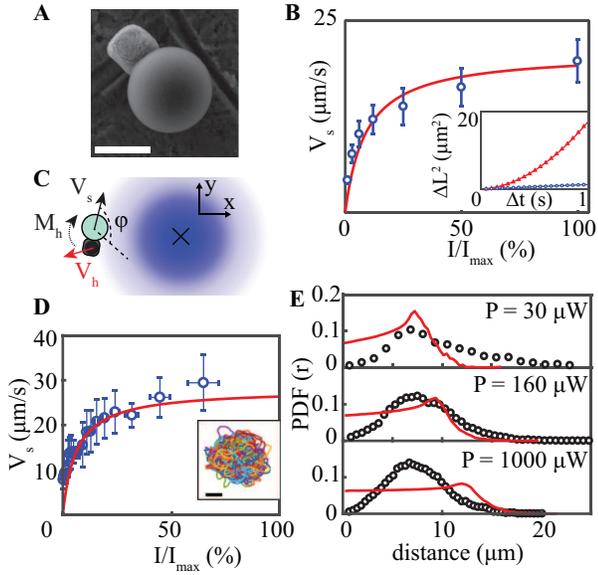}
\caption{\textbf{A phototactic microswimmer}. A) SEM image of the phototactic swimmer, which consists of a haematite particle extruded from a colloidal bead. Scale bar is 1 $\mu$m. B) The swimmer exhibits persistent random walk under homogeneous illumination with a velocity (blue circles) tuned by the light intensity ($I_\text{max} = 550$ nW/$\mu$m$^2$). The red line is a fit using eq.\eqref{eq:swimmer_speed}, and $I_\text{sat}=50$ nW/$\mu$m$^2$. Inset shows the Mean Square Displacement ${(\Delta L)}^2$ of swimmers at equilibrium (no illumination, blue circles) and under illumination (red triangles), at $I/I_\text{max} = 10^{-2}$. Solid lines are fits with eq.\eqref{eq:MSD1}. For $I=0$, we get $D_s=0.3 \pm 0.1$ $\mu$m$^2$/s. C) Scheme of the phototactic mechanism. As the haematite is phoretically repelled from higher intensities of light, it exerts a torque $M_\text{h}$ on the composite swimmer, which orients itself along the gradient. The blue spot represents the laser used in the experiment and the notations are the ones used in the main text to quantify the phototaxis. D) Evolution of the speed $V_s(I)$ of individual swimmers under the same excitation profile and intensities as for the haematite in section \ref{sec:haematite},  ($I_\text{max} =1200$ nW/$\mu$m$^2$). The solid line follows  eq.\eqref{eq:swimmer_speed}, with $V_\text{sat}=24$ $\mu$m/s and $I_\text{sat} = 75$ nW/$\mu$m$^2$. Inset shows the trajectories of the swimmers around the laser spot. Scale bar is 5 $\mu$m. E) Density of Probability of Presence of the swimmers (PDF) at distance $r$ from the center for 3 different excitation powers (black circles). It shows a reasonable agreement with our simulations of ABP , notably the range of the trajectories, without any adjustable parameter (solid red lines, see main text).}
\label{fig:swimmers}
\end{figure}

We analyze the behavior of the swimmers under uniform light and record their trajectories for various intensities (Fig.\ref{fig:swimmers}B). The swimmers self-propel with the bead heading, in line with our previous observations that beads migrate toward the high concentration of \HP\, by  diffusiophoresis\cite{Golestanian_Adjari-PRL-2005,Buttinoni_Bechinger-JPCM-2012,Howse_Golestanian_PRL-2007,Golestanian_Adjari-NewJournalofPhysics-2007,Golestanian_Adjari-PRL-2005,Aubret_Palacci-COCIS-2017,Anderson-AnnualReviewFM-1989} (Fig.\ref{fig:haematite}B). The swimmers exhibit persistent random walk : they travel in straight lines before thermal fluctuations randomize the direction of the motion on a timescale $\tau_r$, the rotational diffusion time .
The Mean Square Displacement (MSD) ${(\Delta L)}^2$ at time $\Delta t$ is extracted as \cite{Howse_Golestanian_PRL-2007,Palacci_Bocquet-PRL-2010_1}:
\begin{equation}
\label{eq:MSD1}
{(\Delta L)}^2 = 4D_s\Delta t + \frac{V_s^2\tau_r^2}{2}\left[\frac{2\Delta t}{\tau_r} + \text{e}^{-2\Delta t /\tau_r}-1\right]
\end{equation}
with $D_s$ the diffusion coefficient of the swimmers at equilibrium (no illumination), and $V_s$ their speed. We measure $D_s$ from equilibrium experiments with 289 swimmers, and find a value of $D_s=0.3 \pm 0.1$ $\mu$m$^2$/s. Following, we extract the rotational diffusion $D_r=3D_s/4R^2=0.3 \pm 0.1$ rad$^2$/s, and the corresponding rotational diffusion time $\tau_r=1/D_r \approx 3$ s. 
We adjust the MSD curves with eq.\eqref{eq:MSD1}, using $V_s$ as the only fitting parameter and obtain the dependency of the phoretic speed on the intensity (Fig.\ref{fig:swimmers}B). The velocity increases rapidly at low intensities, and saturates at higher intensities:
\begin{equation}
V_s(I) = V_\text{sat} \frac{ I}{1+I/I_\text{sat}},
\label{eq:swimmer_speed}
\end{equation}
where $I_\text{sat}$ and $V_\text{sat}$ are the illumination and speed at saturation, respectively. A good agreement is found with the experiment for $V_\text{sat}=21$ $\mu$m/s and $I_\text{sat}\sim 50$ nW/$\mu$m$^2$ at $\lambda = 395$ nm (see Methods), which is consistent with the saturation value for the illumination extracted for the haematite alone, $I_\text{sat}^\text{h}\sim 75$ nW/$\mu$m$^2$ , at a slightly different wavelength $\lambda = 404$ nm.

\subsection{Microswimmer in light gradients}
The behavior of the swimmer qualitatively changes in non-uniform light, as they  migrate towards high intensities  for any given intensity profile (Fig.\ref{fig:swimmers}C and supplementary movie S2), a dynamics akin to the  phenomenon of  phototaxis. \\
In order to quantify the phototactic response, we record the motion of the  microswimmers using the same excitation profiles (intensity and spatial distribution) that we used for the haematite alone in section \ref{sec:haematite} and achieved  by shining a laser spot in the absence of any uniform background illumination. The swimmers travel radially, cross near the center of the laser spot and reverse direction to reorient along the light gradient, towards the center. Note that a similar motion has already been observed for thermophoretic particles oriented by scattering forces around a high intensity laser spot \cite{Moyses_Grier-SoftMatter-2016}.\\
Here, we understand the behavior as a result of the torque exerted on the rest of the swimmer by the haematite, which phoretically migrates towards low intensity of lights (see Fig.\ref{fig:swimmers}C). To test this hypothesis, we quantitatively compare the observed trajectories with a model of a self-propelled particle experiencing a torque from its back part (the haematite), and using measurements independently obtained for swimmers in uniform light and haematite particles alone.
We first reconstruct the trajectories to extract the speed $V_s(r)$ of the swimmers at distance $r$ from the center of the laser spot, and subsequently obtain the intensity dependence $V_s(I)$, under the specific excitation profile (Fig.\ref{fig:swimmers}D). The data for the speed of a microswimmer follows eq.\eqref{eq:swimmer_speed}, with $I_\text{sat} \sim 75$ nW/$\mu$m$^2$, and $V_\text{sat} = 24$ $\mu$m/s. We estimate the orienting torque acting on the swimmer from the velocity of the haematite and the parameter $\chi$ of eq.\eqref{eq:haematite_speed}. \\
Using the values we measured for $\chi$, $I_\text{sat}=75$ nW/$\mu$m$^2$, and $V_\text{sat}=24$ $\mu$m/s, we simulate the trajectories of the swimmers for each investigated intensities with a standard model of Active Brownian Particle (ABP) \cite{Zottl_Stark-JPCM-2016,Marchetti_Simha-RevModPhy-2013}.  
We consider the instantaneous velocity of the swimmer depending only on the local value of the intensity : $V_s =V_s(x,y)$. A swimmer at time $t$ is situated at the position $(x,y)$, with orientation $\mathbf{u_\varphi}$ forming the angle $\varphi$ with the x-axis, in the reference frame centered at $(0,0)$ on the laser spot (Fig.\ref{fig:swimmers}C). The haematite component generates a torque $M_h$ on the swimmer, which pivotes around its vertical axis:
\begin{equation} \label{eq:torque}
\mathbf{M_h}=-R \gamma \mathbf{u_\varphi} \times \mathbf{V_\text{h}} \\
\end{equation}
where $\gamma = 6\pi R \eta$ is the viscous drag coefficient ($\eta$ the dynamic viscosity of the water). The velocity of the haematite alone is radial and writes $\mathbf{V_\text{h}}=V_\text{h}\mathbf{u_r}$, and is given by eq.\eqref{eq:haematite_speed} for a specific intensity profile. We write the Langevin equations for the motion of the swimmer in the $(XY)$ plane as:
\begin{subequations} \label{eq:motion_swimmer}
\begin{align}
& \frac{\text{d}x}{\text{d}t}=V_s(x,y)\mathbf{u_x}\cdot \mathbf{u_\varphi} + \sqrt{2 D_s} \zeta_x(t)\\
& \frac{\text{d}y}{\text{d}t}=V_s(x,y)\mathbf{u_y}\cdot \mathbf{u_\varphi} + \sqrt{2 D_s} \zeta_y(t)\\
& \frac{\text{d}\phi}{\text{d}t}=\frac{M_h}{\gamma R^2} + \sqrt{2 D_r} \zeta_\varphi(t)
\end{align}
\end{subequations}
with $\zeta$ a noise term with zero mean and variance, such that $<\zeta(\tau)\zeta(\tau+t)> = \delta(t)$.
We plot the radial density of probability of presence of the swimmer in Fig.\ref{fig:swimmers}E, and find a good qualitative agreement with the experimental data (Fig.\ref{fig:swimmers}E). In the absence of any fitting parameters, we notably reproduce the spatial extension of the trajectories around the laser spot. Discrepancies between the simulations and the experiment arise at short distance $r$, where the focus of the laser is intense ($I\sim \mu$W/$\mu$m$^2$). Scattering of light by haematite, repelled from the center by radiative pressure and  \HP\, decomposition due to the UV light could both lead to repulsion away from the laser sport, which we did not attempt to capture in our model. It offers  a simple route to design self-propelled particles whose motion is controlled by light patterns.

\begin{figure*}[th!]
\centering
\includegraphics[scale=1]{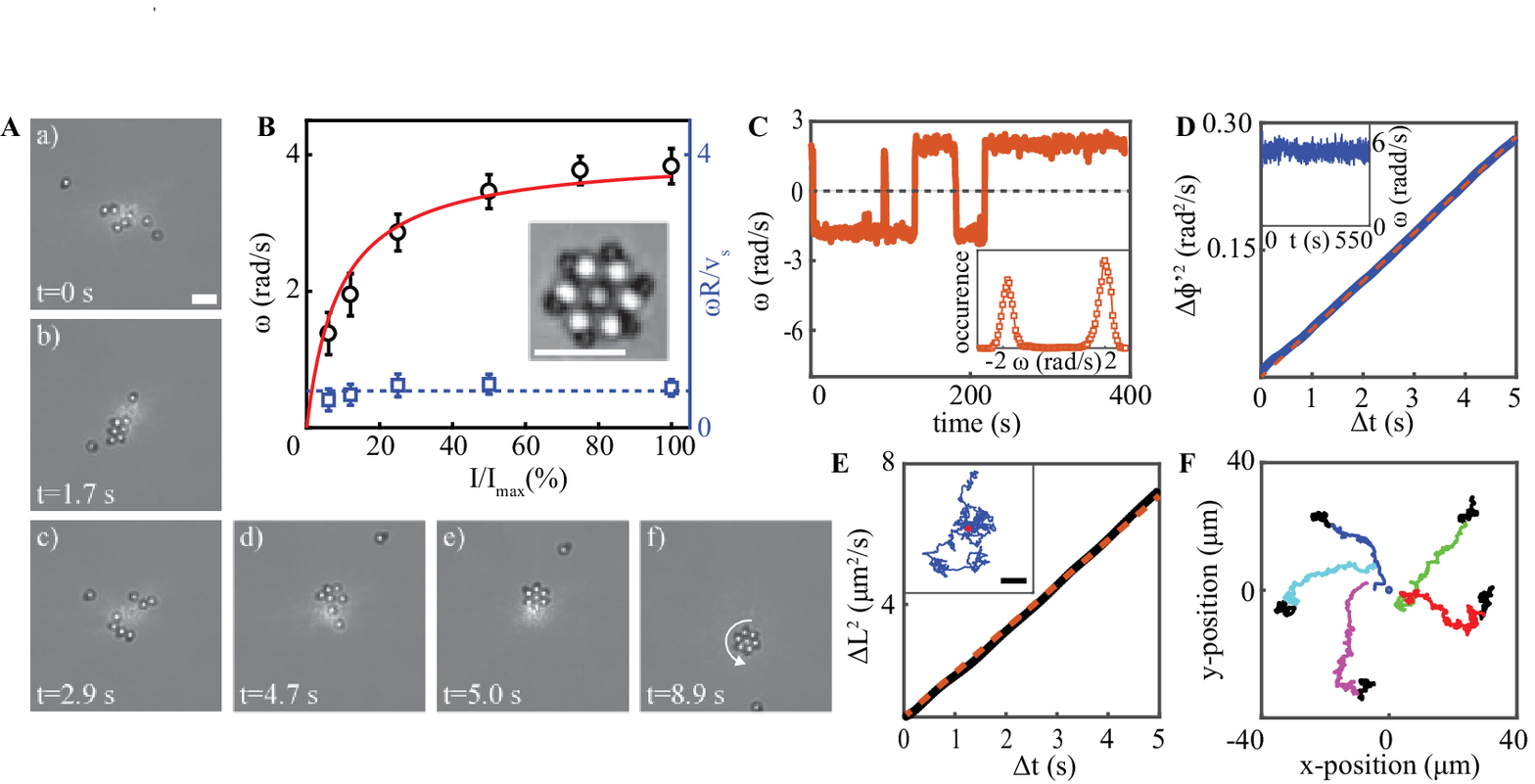}
\caption{\textbf{Formation mechanism and properties of the rotor}. A) Snapshots of a movie showing the assembly of a rotor around a single laser spot. In (a) the swimmers travel around the laser spot, until they assemble, (b) and form a transient structure that breaks,  (c). The swimmers collide again, (d) and form a stable rotor (e), which is sustained under illumination after removing the gradient (f). B) The angular speed of the rotor is tuned by the light intensity. The solid red line is an adjustment with eq.\eqref{eq:swimmer_speed}, using the same saturation intensity as for the swimmers alone. The angular speed reflects the speed of the swimmers (blue squares). Inset: magnified image of a rotor, where the orientation of the swimmers is visible. C) At low angular speeds ($\omega < 3$ rad/s), the rotor can flip its direction of rotation. Inset: distribution of speeds for a rotor during a $\sim 10$ min experiment. The two peaks are situated at -1.8 and 1.9 rad/s, showing the symmetry of the dynamics. D) At higher speeds, the motion is persistent and the angular fluctuations can be extracted (blue line). Dashed red line shows a linear adjustment using eq.\eqref{eq:MSD2}, and gives $D_\Phi =0.03 \pm 0.01$ rad$^2$/s. Inset: speed of a fast rotor ($\omega \sim 6$ rad/s) over 10 min. E) Experimental MSD (black curve) of free-to-move rotors, and linear adjustment (red dashed line), giving $D_\text{R} = 0.4 \pm 0.1$ $\mu$m$^2$/s. Inset: trajectory of the center of mass of rotor optically trapped with a red laser (red) and free to diffuse (blue) (see main text and Methods).  F) Migration of rotors in light gradient, for 5 different experiments.  The rotors first exhibit a diffusive motion in homogeneous light, (black curves, 15 s for each experiment) but migrate towards the bright spot (colored curves) of a focused laser light (circle at the center). Typical time for migration is $10 - 30$ s. All scale bars are 5 $\mu$m. }
\label{fig:rotorsproperties}
\end{figure*}

\section{Guided self-assembly using light patterns}
In this section,  we describe the use of light patterns to guide and trigger the self-assembly of microswimmers  into self-spinning microgears (see supplementary movie S3). 
We start from a dilute sample of phototactic microswimmers under uniform illumination. They travel along the substrate and the density $\Phi_s\sim 10^{-3}$ part/$\mu$m$^2$ is too low to observe formation of clusters by Motility Induced Phase Separation (MIPS) \cite{Stenhammar_Cates-PRL-2013,Cates_Tailleur-ARCMP-2015}. Following, we superimpose a focused laser spot to locally increase the density of particle and trigger collision. Responding to the light gradient, the swimmers direct their motion towards the laser spot, where they meet, collide, and eventually form a self-spinning microgear. It is composed of 7 particles, consisting of one central swimmer with the haematite pointing up, surrounded by 6 close-packed particles facing the central one. The peripheral swimmers collectively orient in the same direction, setting chirality and  rotation of the assembly. Once the structure is formed, the laser can be switched off and the microgear is stable and spins under homogeneous illumination (see Fig.\ref{fig:rotorsproperties}A). 

\subsection{Microgears self-assembly}
Haematite scatters in the blue leading to shadowing\cite{Chen_Zhang-AdvancedMaterials-2017} that sets a vertical gradient of \HP. This effect is usually negligible but becomes significant at higher intensity of  light when it can produce phoretic lifting. Subsequently, as the particle meet and collide close to the laser spot, one of the microswimmer is flipped as the haematite component is phoretically lifted up by the laser light. The swimmer sits vertical, facing the substrate and pushing against the wall of the bottom substrate. It induces a pumping flow\cite{DiLeonardo_Ruocco-Langmuir-2009,Lopez_Lauga-PhysicsofFluid-2014}, that attracts neighboring swimmers into hexagonal closed-packing. After formation, the structure is stable and the  laser can be switched off. The rotors remain as long as the uniform illumination of the background allow the particles to consume fuel. Each haematite part of the microswimmers in the structure constitutes a chemical sink of \HP\, that interacts with the neighboring swimmers (see section 2). The diffusiophoretic repulsion between the polymer spheres and the haematite components of the peripheral swimmers induces the collective tilt and orientation of the particles in the shell, setting the chirality and the spinning of the structure. Moreover, the repulsion prevents the formation of an second layer of peripheral swimmers and additional particles spontaneously detach, warranting the fidelity and robustness of the self-assembly (see supplementary movies S3, S4). 

\subsection{Rotors dynamics}
We start by forming isolated rotors , superimposing a laser spot to a uniform illumination background, with $P >1$ mW the minimal power of the laser (spot size of 0.5 - 5 $\mu$m) to trigger the self-assembly. Once the rotor is formed, we track the motion  of all particles forming the microgear, and reconstruct their individual trajectories. The experiment is repeated for multiple independent rotors and intensity of light  to study the dynamics. \\
The center of mass of individual rotors exhibits a diffusive behavior in uniform light, with diffusion $D_\text{R} = 0.4 \pm 0.1$ $\mu$m$^2$/s (Fig.\ref{fig:rotorsproperties}E). This value is higher than the diffusion constant expected from a passive particle of this size ($R\approx 3$ $\mu$m) and arises from active noise. The center of the rotors  can moreover be optically trapped by a low intensity red laser ($\lambda = 685$ nm, $P\sim 10-100$ $\mu$W) to suppress the random motion in the center of mass (Fig.\ref{fig:rotorsproperties}E-Inset), without further disturbing the system as haematite is weakly absorbing  in the red wavelength, which has negligible effect on the photocatalytic activity. 
The angular speed of the rotation is tuned by the light intensity and reflects the change of translational speed of the swimmer $V_s$ with a $\sim \pi/4$ incline: $\omega R/V_s \sim \sqrt{2}/2$ (Fig.\ref{fig:rotorsproperties}B). At low rotation rate ($\omega< 3$ rad/s), thermal fluctuations can flip the direction of the spinning, but the magnitude of the speed of the rotor remains constant (Fig.\ref{fig:rotorsproperties}C). Further reducing the intensity reduces the stability of the rotor, which breaks  back into 7 individual microswimmers. At higher speeds ($\omega>3$ rad/s, Fig.\ref{fig:rotorsproperties}D-Inset), the rotors are robust and the motion is persistent over the duration of the experiment (>20 min). The handedness is random, and we measure 49.85\% clockwise and 50.15\% of counter clockwise for 1017 rotors. \\ 
We  quantify and analyze the angular dynamics of the rotors. Assuming a $\delta$ correlated gaussian noise $\zeta(t)$, the phase $\Phi$ follows the Langevin equation \cite{Wykes_Shelley-SoftMatter-2016,Rosenblum_Pikovsky-ContemporaryPhysics-2003} :
\begin{equation}
\label{eq:MSD2}
\frac{\text{d}\Phi(t)}{\text{d} t} = \omega +\sqrt{2D_\Phi}\zeta(t)
\end{equation}
with $\omega$ the mean angular speed, and $D_\Phi$ the rotational diffusion coefficient of the rotors.
From this, we compute the angular fluctuations $\Delta \Phi'^2 = 2D_\Phi t$ with $\Phi'=(\Phi-\omega t)$ by analyzing the angular dynamics of 12 individual rotors with the speed $\omega$ extracted as the mean value of the angular speed. The ensemble average is linear in time, as  shown in Fig.\ref{fig:rotorsproperties}D and gives an amplitude for the angular noise, $D_\Phi = 0.03 \pm 0.01$ rad$^2$/s, comparable with the translational diffusion measured for the center of mass of the rotors, stressing their common active origin. Linking the translational and angular active noise as given by the Stokes Einstein relation in an equilibrium system, we find that the temperature of the active bath reaches  $T\approx 1700$K\cite{Palacci_Bocquet-PRL-2010_1}.\\
We further use the blue laser to generate light gradients and manipulate rotors (see Fig.\ref{fig:rotorsproperties}F). First, the rotors assemble and spin under  uniform light.  Then, we switch on the blue laser to superimpose a light gradient. The swimmers inside the rotors reorient by phototaxis without breaking the microgear. A slow migration at $\sim 1 $ $\mu$m/s towards the laser spot follows, without breaking of the rotor, showing their robustness under manipulation.\\
The overall properties of the  rotors are  all summarized in Fig.\ref{fig:rotorsproperties}. As the rotors consume fuel and alter their surrounding, we are now interested in measuring and quantifying the dissipative diffusiophoretic interactions. 
 
\section{Anisotropic and dynamic diffusiophoretic interactions}
We model our system to quantify the strength and shape of the diffusiophoretic interactions around a rotor and compare with  experimental measures. To this end, we use  HILO microscopy to image and quantify the effective interaction created by a rotor on neighboring particles. We stress that the system is in the so-called  low P\'{e}clet regime $Pe \sim \omega R^2/D \ll 1$ as the diffusion of chemicals, with diffusivity $D$, is fast compared to the rotation of the rotors. It allows us to consider the equilibration of the cloud of chemicals as {\it instantaneous} and the concentration field of \HP\,   steady in the rotating frame of the rotor.

\subsection{Analytical solution}
We first derive an analytical expression considering a simplified model. We model the rotor by a sphere, with a surface activity $\nu(r=\mathbf{R_s},\theta,\phi)$ in spherical coordinates. The modulation of surface activity $\nu$ reflects the positioning of the haematite on the periphery of the rotor.
The concentration $c$ of \HP\, follows\cite{Golestanian_Adjari-NewJournalofPhysics-2007}:
\begin{subequations} 
\begin{align}
& D\nabla^2c=0;\\
& -D\mathbf{n}\cdot\nabla c(\mathbf{R_s}) = \nu(\mathbf{R_s})
\end{align}
\label{eq:Laplace}
\end{subequations}
We expand the concentration on a basis of spherical harmonics $Y_{l}^m (\theta,\phi)$. For a surface activity of the form $\nu(\theta,\phi) =  \sum\limits_{l=0}^{\infty} \sum\limits_{m=-l}^{m=+l} \nu_{lm} Y_{l}^m(\theta,\phi)$, we get from  \eqref{eq:Laplace}:
\begin{equation}
c(r,\theta,\phi) = c_\infty + \frac{R}{D} \sum_{l=0}^{\infty} \sum\limits_{m=-l}^{m=+l} \frac{\nu_{lm}}{l+1} {\left(\frac{R}{r}\right)}^{l+1} Y_{l}^m (\theta,\phi)
\label{eq:concentration}
\end{equation}

\begin{figure*}[th!]
\centering
\includegraphics[scale=1]{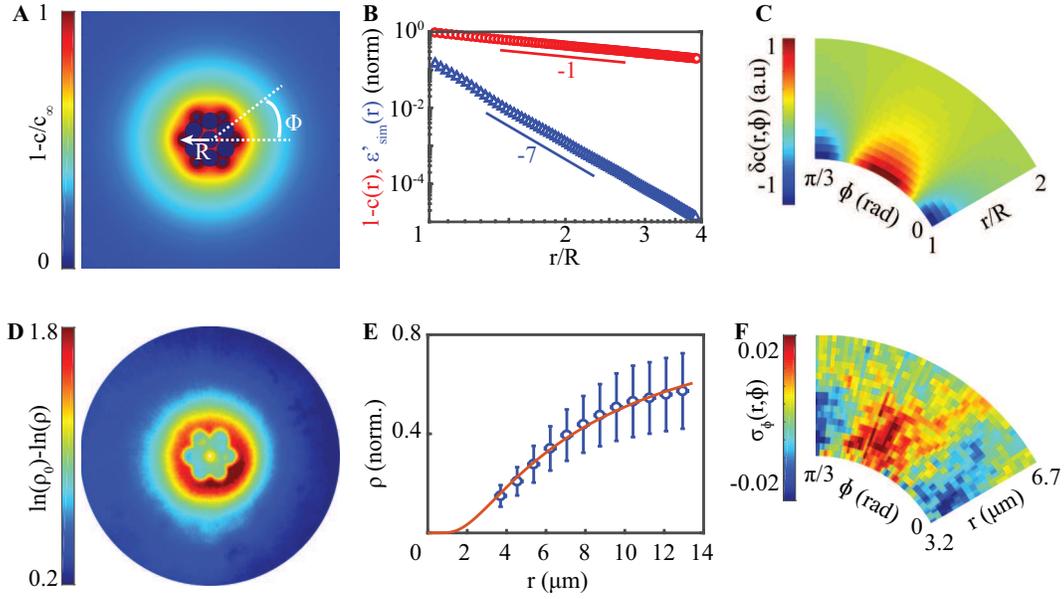}
\caption{\textbf{Anisotropic and diffusiophoretic interactions}. A), B), C) are simulations and D), E), F) are experiments. A) Numerical mapping of the (normalized) concentration field around a rotor of radius $R$, situated above an impermeable interface. At fixed distance, the concentration profile is periodically modulated by the presence of the haematite. B) Evolution of the radial concentration profile (after azimuthal average, red circles), following a $\propto 1/r$ decay. The azimuthal profile follows a sinusoidal variation in the interval $[0;\pi/3]$ with decaying amplitude $\epsilon'_\text{sim}\propto 1/r^7$ (blue triangles, see main text). C) Mapping of the azimuthal variations after subtracting the radial dependency (see main text). D) Experimental observation of the interaction profile between a rotor and 20 nm fluorescent beads using HILO microscopy, in the reference frame of the rotor. The data closely resemble A). E) The radial density profile of the fluorescent beads (blue circles) follows eq.\eqref{eq:density_radial} (red line), with $\alpha = 120$ $\mu$m$^3$/s. F) Experimental mapping $\sigma_\phi$ of the azimuthal interactions after removal of the radial contribution (see main text). A periodic modulation is visible, with the higher repulsion near the haematite.}
\label{fig:repulsion}
\end{figure*}

The  decay of the concentration is given by the multipolar order of the surface activity. We prescribe a surface activity $\nu(\theta, \phi) =\nu_0 (1-\sin^6{\theta} \cos{6\phi})$, where the hexapolar symmetry of the rotor  is given by  $\cos{6\phi}$ term and $\sin^6{\theta}$ confines the activity to the plane of the rotor. It conveniently gives $\nu \propto Y_{6}^6(\theta,\phi) +Y_{6}^{-6}(\theta,\phi)$, such that only ${l=0,  6}$ remain in the series expansion of eq.\eqref{eq:concentration} so that:
\begin{equation}
c(r,\theta,\phi) = c_\infty -\frac{\nu_0 R^2}{Dr} - \frac{\nu_0 R^8}{7Dr^7}\cos{6\phi} \sin^6{\theta}
\label{eq:concentration_final}
\end{equation}
which we write
\begin{equation}
c(r,\theta,\phi) = c_\infty -\alpha'(r) - \epsilon'(r)\cos{6\phi} \sin^6{\theta}
\label{eq:concentration_final2}
\end{equation}
with $\alpha' = \nu_0 R^2/Dr$ and $\epsilon' =  \nu_0 R^8/7Dr^7$. The parameter $\alpha'$ represents the amplitude of the long-range radial diffusiophoretic interaction, while the strength of the azimuthal interaction is given by $\epsilon'$. They can be experimentally extracted by measuring the density profile of fluorescent beads around a rotor, and according to the model, their ratio is $\epsilon' r^6/\alpha' R^6 =1/7$.
Using particles to probe the diffusiophoretic interactions, as we will do in the experiment (see sec. 5.3), the conservation of number of probes follows eq.\eqref{eq:flux} at $(r,\theta,\phi)$. At steady state, $\mathbf{J} = 0$, it leads  to the following expression for the density $\rho$ of probe particles:
\begin{equation}
\rho(r,\phi,\theta) = \rho_0 \exp{(-\alpha' - \epsilon'\cos{6\phi} \sin^6{\theta})}
\label{eq:densityprofile_rotor}
\end{equation}

For later convenience, we define:
\begin{equation}
\sigma_\phi (r,\phi)= \ln{\rho(r,\phi)} - \langle \ln{\rho(r,\phi)} \rangle_\phi = \epsilon'\cos{6\phi}
\label{eq:sigma}
\end{equation}
for $\theta = \pi/2$ (XY plane of the rotor). This quantity will serve to experimentally evaluate $\epsilon'(r) \propto 1/r^7$, as we shall see.

\subsection{Numerical simulations}
\label{sec:rotors_sim}
We numerically solve the steady diffusion equation of the  \HP\, concentration $\Delta c=0$ around a rotor in 3D (COMSOL Multiphysics 5.3).  We model the rotors as 7 passive and impermeable spheres (one central and six peripheral) of radius $R_c$, decorated at the periphery by 6 chemically active core-shell spheres of radius $R_c/2$, plus one at the top of the central swimmer. The fuel consumption is imposed by a constant reaction rate occurring in a thin shell of $R_c/10$. The cores of the active particles and the passive spheres are set as impermeable, with no-flux boundary condition. We position the rotor at distance $R_c/10$ above an (infinite) impermeable wall constituted by the bottom surface of the capillary in the experiment. The concentration is set at 1 at infinity in upper space using Infinite Element Domain (COMSOL) to simulate infinite boundary conditions in a spherical coordinate system. \\	
We first extract the concentration profile in the radial direction after azimuthal averaging in the XY plane, and measure a $1/r$ decay (see Fig.\ref{fig:repulsion}A-B). We then extract the azimuthal dependence of the concentration and calculate $\delta c(r,\phi) = c(r,\theta=\pi/2,\phi)-\langle c(r,\theta=\pi/2,\phi) \rangle_\phi$.
For each radial distance $r$, $\delta c (\phi)$ curve show a cosine shape (Fig.\ref{fig:repulsion}C), from which we extract the amplitude $\epsilon'_\text{sim}(r) =\left[\delta c_\text{max} - \delta c_\text{min}\right]/2$ . The evolution of $\epsilon'_\text{sim}(r)$ is shown in Fig.\ref{fig:repulsion}B,  where it decays as $1/r^7$. In particular, we extract the ratio $\epsilon'(r) r^6/\alpha'R^6 = 0.18 \pm 0.04$ for $1<r/R<4$, in excellent agreement with the analytical model and eq.\eqref{eq:concentration_final}, which predicts $\epsilon' r^6/\alpha' R^6=1/7\approx 0.14$.

\subsection{Optical imaging of the concentration field}
We aim at comparing our predictions for the diffusiophoretic interactions with experimental measurements.
We visualize the effective interaction around a single rotor by dispersing  20 nm nile red fluorescent beads in the solution that exhibit  diffusiophoretic motion in \HP\, gradients  (Fig.\ref{fig:repulsion}D and setup description in the Methods section). We first form a rotor, that we optically trap with a red laser to {\it pin} its center and suppress the random motion of its center of mass (see Methods). We lower the intensity of the illumination to slow down the dynamics of the rotor and let the time of the fluorescent beads to equilibrate, in order to image the -steady- density profile in the reference frame of the rotor. We use a home-built HILO microscopy setup (see Methods) to visualize a slice of a few $\mu$m above the substrate and suppress the bulk background fluorescence (see Methods). Care was taken to  carefully align the trapping laser with the center of the HILO illumination, where the depth of the excitation field is minimal. We then excite the beads and record their fluorescence during 1 s time intervals, imposing dark periods of $4$s between successive excitation cycles. This prevents strong photo bleaching of the area around the rotor, as the small beads have time to diffuse away during the dark periods. The sequence is repeated for 10 cycles, or when the rotor breaks. We record the background fluorescence in independent experiments under the same excitation conditions  and correct accordingly.
\begin{figure}[th!]
\centering
\includegraphics[scale=1]{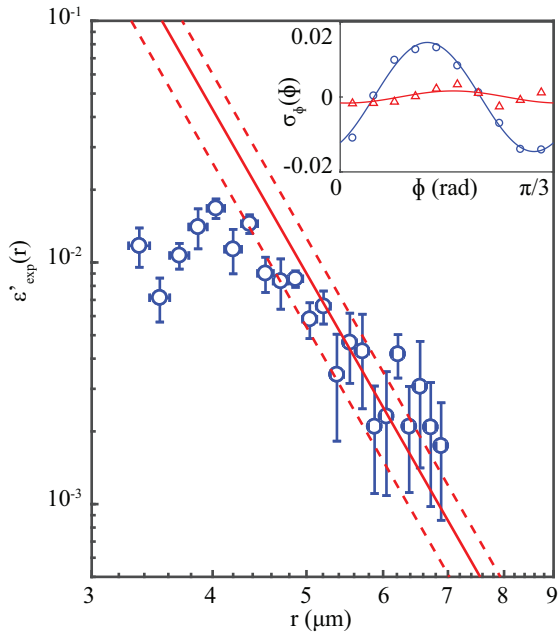}
\caption{\textbf{Experimental determination of the azimuthal interaction.} Evolution of the amplitude of oscillations shown in Fig.\ref{fig:repulsion}F, as a function of the distance decaying as $\propto 1/r^7$. Red solid and dashed lines are the mean value and its extrema, respectively, as predicted by the  numerical simulation in the absence of adjustable parameter (see main text). It shows an excellent agreement with the experimental data. Inset: Profiles of Fig.\ref{fig:repulsion}F, at distances of 4 (blue circles) and 6 $\mu$m (red triangles) from the center of the rotor. Solid lines are adjustments using a sinusoidal function. It confirms diffusiophoresis as the source of interaction.}
\label{fig:rotor_epsilon}
\end{figure}
We analyze the movies by measuring the instantaneous fluorescence profile around a rotor in the rotating frame $(r,\phi)$. Averaging over all the frames, we obtain the intensity profile (Fig.\ref{fig:repulsion}D), further averaged  over $\phi \in \left[0;\pi/3\right]$ using the symmetry of the rotor.\\
In the far field, the concentration of fluorescent particles follows $\rho=\rho_0 e^{-\alpha/(D_\text{c}r)}$ as prescribed in eq.\eqref{eq:density_radial} for a sink of \HP, $c\propto 1/r$ (Fig.\ref{fig:repulsion}E). We extract the amplitude of the radial phoretic repulsion $\alpha = 120 \pm 30$ $\mu$m$^3$/s, in line with the results from the experiments realized with haematite alone and the same fluorescent beads (section \ref{sec:haematite}).
In the near field,  1<r/R<2, the concentration field  is hexapolar and we  investigate the azimuthal gradient of concentration by computing the experimental values for $\sigma_\phi=\ln{\rho(r,\phi)}-\langle{\ln{\rho(r,\phi)}}\rangle_\phi$. We observe a small modulation of the amplitude in the interval $[0;\pi/3]$, for distances $r\sim 3.2 - 6.7\mu$m (Fig.\ref{fig:repulsion}F), with the maximal repulsion  near the haematite. Guided by the analytical model [eq.\eqref{eq:sigma}], we adjust the data with the function $\epsilon'_\text{exp}\cos(6\phi+\phi_0)$ using $\epsilon'_\text{exp}$ and $\phi_0$ as fitting parameters and repeat the procedure for different distance $r$ to the center of the rotor, giving $\epsilon'_\text{exp}\propto 1/r^7$  (Fig.\ref{fig:rotor_epsilon}), as predicted by eq.\eqref{eq:densityprofile_rotor}. We  further compare the experimental value with the estimate from the  numerical simulations using  $\epsilon'_\text{sim}(r) r^6/\alpha'R^6 = 0.18 \pm 0.04$ (section \ref{sec:rotors_sim}) and $\alpha = 120$ $\mu$m$^3$/s, showing  an excellent agreement without adjustable parameters (Fig.\ref{fig:rotor_epsilon}). Further, those interactions can be used to synchronize multiple rotors and in fact, the results hereby described with probe particles agree with the  phoretic coupling parameter extracted in ref.\cite{Aubret_Palacci-NaturePhysics-2018} for interacting rotors. They confirm diffusiophoresis as a powerful tool to  shape and program interactions with a quantitative control.  

\section{Conclusion}
We demonstrate the successful building of hierarchical structures from a simple type of dissipative building block. To this end we harnessed diffusiophoretic interactions to guide the self-assembly and encode the dynamics. Starting from a photoactive material, the haematite, we designe phototactic microswimmer that migrate toward high intensity of light. They spontaneously assemble into self-spinning microgears, that   rotate and sustain in the absence of any feedback from the operator. 
We introduce an optical method - HILO microscopy - as a novel tool to characterize phoretic interactions. The method is generic and can be implemented to image phoretic interactions around (self)-phoretic swimmers. It reveals anisotropic, periodic near field modulations of the chemical cloud around rotors, which amplitude compares with both numerical simulations and a simple analytical model.
Our rational design of machines from machines shows the opportunities offered by phoretic phenomena to control dissipative self-assembly, and paves the way to further study collective effects in large assemblies of spinning structures.

\section{Materials and Methods}

\label{sec:setup1}
\begin{figure*}[th!]
\centering
\includegraphics[scale=1]{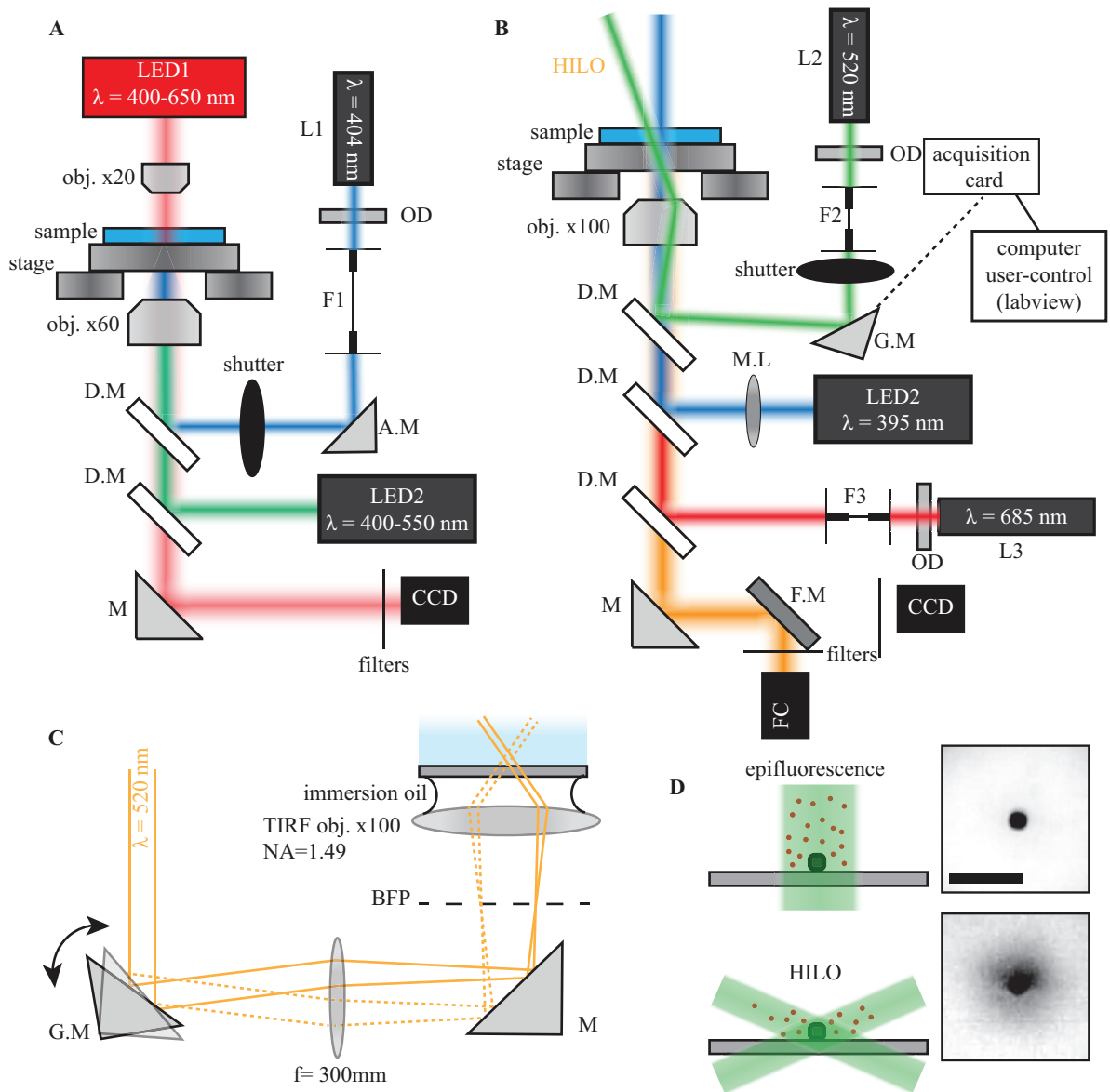}
\caption{\textbf{Experimental setups}. See main text for full description. A) Experimental setup used for rotor creation. B) Highly Inclined Laminated Optical sheets (HILO) microscopy setup used for imaging the concentration profile around a rotor. C) Detailed scheme showing the light path for HILO microscopy. The laser is focused at the back focal plane of a high-NA objective. Shifting the position of the laser with respect to the center of the entrance pupil results in inclined, output light beams. D) \textit{Top}:In standard epifluorescence microscopy, background fluorescence is high because of strong illumination of the planes above the focal plane. \textit{Bottom}:In HILO microscopy, only a small slice ($\sim$ few microns) above the focal plane is shined, lowering the background fluorescence. The images on the right show the fluorescence profile of beads around a haematite in epifluorescence (top), compared to HILO (bottom). Scale bar is 5 $\mu$m. The symbols used are : \textit{L1, L2, L3} : Laser sources; \textit{D.M} : Dichroic Mirror; \textit{G.M} : 2-axis Galvanometric Mirror; \textit{F1}: Multimode Fiber; \textit{F2, F3}: Single Mode fibers; \textit{OD}: Optical Densities; \textit{F.M} : Flip Mirror; \textit{A.M} : Adjustable Mirror \textit{M.L} : Movable Lens; \textit{F.C} : Fluorescence Camera; \textit{CCD}: Charged-Coupled-Device camera; \textit{M}: Mirror. For better clarity, we do not represent optical lenses.}
\label{fig:setup}
\end{figure*}

\subsection{Sample preparation}
The samples are prepared at low particle density $\Phi_s \sim 10^{-3}$ part/$\mu$m$^2$.  Particles are diluted in a 6$\%$ solution of hydrogen peroxide H$_2$O$_2$ (Fisher Scientific H325-500) in deionized water (Milli-Q, resistivity 18.2 M$\Omega$). The solution is then injected in a small rectangular capillary (VitroCom 3520-050), and sealed with capillary wax (Hampton Research HR4-328). All the capillaries are previously subjected to plasma-cleaning (Harrick Plasma PDC-001) and rinsed thoroughly with deionized water. As the particles are non-buoyant, they sediment near the bottom surface of the capillary, and observation is made in this plane on an home-built microscope.\\
Synthesis of the particles is described in details in the supplementary information of ref.\cite{Aubret_Palacci-NaturePhysics-2018}.

\subsection{Measurements of intensities}
-\,To extract the intensity profile of the blue laser in sections \ref{sec:haematite} and \ref{sec:swimmers}, we deposit fluorescent beads (Lifetechnologies F8888, diameter 20 nm) on the surface of a capillary, and fill it with water to obtain the fluorescence intensity pattern along the radial direction after azimuthal averaging.\\
-\,The maximum power from LED2 arriving on the sample is measured to be of the order of $\sim 10$ mW, on an area of roughly $150 \times 150$ $\mu$m$^2$. This corresponds to an intensity $I_\text{max}$ of $\sim 500$ nW/$\mu$m$^2$. We find $I_\text{sat}/I_\text{max}=9\%$ and thus $I_\text{sat} \sim 50$ nW/$\mu$m$^2$.\\
-\,For the blue laser, the shape of the intensity profile in Fig.\ref{fig:haematite}C is used to compute the peak intensity for a 1 mW power, and gives $I_\text{peak} = 2000$ nW/$\mu$m$^2$. From the experiments, we extract $I_\text{sat} = 75$ nW/$\mu$m$^2$.\\

\subsection{Optical setup for rotor creation}
The custom optical setup is represented on Fig.\ref{fig:setup}A.
The sample is observed with optical microscopy on an inverted microscope and with bright field transmitted illumination (LED1).
A LED is set up in the blue range (LED2, $\lambda=425-500$ nm, Lumencore SOLA 6-LCR-SC) and uniformly illuminates the sample on a large area (up to $300\times 300$ $\mu$m$^2$) to trigger the photo-catalysis of haematite, after being reflected on an appropriate dichroic mirror. A laser diode source L1 ($\lambda=404$ nm, Thorlabs L404P400M) is superimposed to the homogeneous excitation profile. It is focused in a multimode (MM) optical fiber F1 (Thorlabs M42L02), re-collimated at the output, and reflected on a manually adjustable mirror (A.M). The beam is then reflected on a dichroic mirror (Thorlabs DMLP425R) before being focused on the sample through a high Numerical Aperture (NA) oil-immersion objective (Nikon 60x, NA=1.4). The adjustable mirror allows to steer the laser spot along the surface of the sample and select the location for rotor formation. Furthermore, an electronic shutter (Thorlabs SHB1T) on the optical path enables switching ON and OFF the laser spot.
The sample is mounted on a manual micrometric stage (Nikon Ti-SR). We observe the image of the sample through the 60x objective on a monochrome CCD camera (Edmund Optics EO-1312M), with a set of appropriate optical filters. Acquisitions are performed at typical rates of $20-100$ fps. 

\subsection{Optical setup for HILO microscopy}
\subsubsection{HILO microscopy}
We modified the optical setup presented in section \ref{sec:setup1} to analyze the phoretic repulsion of fluorescent beads around a haematite and a rotor (Fig.\ref{fig:setup}B). We used Highly Inclined and Laminated Optical sheet microscopy (HILO, also called near-TIRF) to visualize the fluorescence in a thin slice ($\sim$ few $\mu$m) above the surface \cite{Tokunaga_Sogawa-NatureMetehods-2008,Shashkova_Leake-BioscienceReports-2017,Liu_Betzig-MolecularCell-2015}. The setup is similar to a Total Internal Reflection Fluorescence Microscopy setup. In HILO microscopy, the beam arrives at an angle slightly smaller than the critical angle, such that the beam is highly inclined, and not totally reflected. This results in the illumination of a thin slice with a width of typically a few microns\cite{Tokunaga_Sogawa-NatureMetehods-2008} (Fig.\ref{fig:setup}C-D) and therefore much lower background fluorescence compared to epifluorescence microscopy, as exemplified in Fig.\ref{fig:setup}D.
\subsubsection{Optical setup}
We used nile red fluorescent beads (F8784, ThermoFisher Scientific) to image the concentration profile, so that excitation at $\sim 520$ nm for fluorescence at $\sim 575$ nm can be done without activating the haematite, weakly absorbing at this wavelength.
The blue laser light is replaced by a $520$ nm laser light source L2 (PL520, Thorlabs, 50 mW) and focused in a single mode fiber F2 (SM, Thorlabs, P1-460B-FC-2) so that the resulting output beam mode is gaussian. The beam is collimated again, and sent in a 2D galvanometric mirror system (G.M, Thorlabs GVS212). Using a long-focal achromatic lens (300 mm), the beam is focused at the back focal plane of a TIRF objective with high numerical aperture (apo-TIRF, x100, NA=1.49, NIKON). The output beam from the objective is collimated, and its angle with the surface of the sample is adjusted by controlling the orientation of the galvanometric mirror, addressed in real time through an acquisition card (NI USB-6343) with a home-made user-controlled interface developed under Labview 2017. In particular we impose a high frequency (100 Hz) oscillatory motion of the mirror to scan over $2\pi$ the entrance pupil of the objective with the laser beam. This technique guarantees an isotropic illumination profile.\\
We form the rotors by changing the wide-field excitation pattern of LED2 using a movable lens (ML). The swimmers migrate toward the high intensity area, so that reducing the size of the light spot increases the density and eventually triggers the formation of the rotor. Adjusting the position of the lens back to its original position disperses the swimmers again in a homogeneous intensity profile and keeps the rotor active.
Finally, a red laser L3 ($\lambda = 685$ nm, Thorlabs, HL6750MG, 50 mW) is added to the setup to trap the rotors. The beam is focused in a single mode fiber F3 (SM, P1-630A-FC-2), recollimated at the output, and reflected on a shortpass dichroic mirror (Thorlabs, DMSP650R). The position of the beam is carefuly adjusted and centered on the HILO excitation pattern.
We observe the fluorescence of the beads on a fluorescence camera (Photometrics Scientific, Coolsnap HQ2), with a resolution of $0.168$ $\mu$m/pixel, and with an integration time of $50$ ms. A flip mirror (F.M) is inserted on the detection path to switch between viewing of the fluorescence or the bright field image on the CCD camera, using sets of appropriate spectral filters for each camera.
The overall setup allows to form and trap a rotor, and to observe its near-field concentration cloud through HILO microscopy.

\section*{Conflicts of interest}
The authors declare no conflicts of interests

\section*{Acknowledgements}
The authors gratefully acknowledge Stefano Sacanna and Mena Youssef from New York University for supplying the colloidal particles. This material is based on work supported by the National Science Foundation under Grant No. DMR-1554724. J.P. thanks the Sloan Foundation for support through grant FG-2017-9392.
%%%END OF MAIN TEXT%%%

\bibliography{BiblioAA} %You need to replace "rsc" on this line with the name of your .bib file
\bibliographystyle{achemso} %the RSC's .bst file

%\balance


%%%REFERENCES%%%

\end{document}